**Title:** Generation of 30-fs pulses from a diode-pumped graphene mode-locked Yb:CaYAlO$_4$ laser

*Jie Ma[1,2], Haitao Huang[1,3], Kaijie Ning[1], Xiaodong Xu[3], Guoqiang Xie[4,5], Liejia Qian[4,5], Kian Ping Loh[2,\*], and Dingyuan Tang[1,5,\*]*

*Corresponding Author: E-mail: chmlohkp@nus.edu.sg, EDYTang@ntu.edu.sg

[1] School of Electrical and Electronic Engineering, Nanyang Technological University, Singapore 639798, Singapore
[2] Department of Chemistry and Graphene Research Centre, National University of Singapore, Singapore 117543, Singapore
[3] School of Physics and Electronic Engineering, Jiangsu Normal University, Xuzhou 221116, China
[4] Key Laboratory for Laser Plasmas (Ministry of Education), Department of Physics and Astronomy, Shanghai Jiao Tong University, Shanghai 200240, China
[5] IFSA Collaborative Innovation Center, Shanghai Jiao Tong University, Shanghai 200240, China

Keywords: graphene, ultrafast lasers, mode-locking, saturable absorbers

Abstract: Stable 30 fs pulses centered at 1068 nm (less than 10 optical cycles) are demonstrated in a diode pumped Yb:CaYAlO$_4$ laser by using high-quality chemical vapor deposited monolayer graphene as the saturable absorber. The mode locked 8.43 optical-cycle pulses have a spectral bandwidth of ~ 50 nm and a pulse repetition frequency of ~ 113.5 MHz. To our knowledge, this is the shortest pulse ever reported for graphene mode-locked lasers and mode-locked Yb-doped bulk lasers. Our experimental results demonstrate that graphene mode locking is a very promising practical technique to generate few-cycle optical pulses directly from a laser oscillator.

## 1. Introduction

Ultrafast solid-state lasers, especially the diode-pumped femtosecond lasers, have attracted considerable attention in recent years due to their high efficiency, compact size, low cost, and promising applications in scientific research and industry. Benefiting from the rapid advance of material researches, various bulk laser gain materials with broad emission spectrum and high thermal conductivity have been developed with the aim of generating sub-50 fs ultrashort pulses. Worth of mentioning that in the 800 nm region few-cycle pulses were successfully obtained from different Ti:sapphire lasers based on either the semiconductor saturable absorber mirrors (SESAMs) [1, 2] or the Kerr-lens mode-locking technique [3-6]. These two mode locking methods are the current mainstream technique for generating ultrashort pulses in mode locked bulk lasers. However, the fabrication of SESAMs requires complex and sophisticated clean-room manufacturing technology [7]. Moreover, SESAMs generally have a limited operation bandwidth (usually no more than 100 nm), which is difficult to support few-cycle mode-locked pulse generation in the infrared region. The Kerr-lens mode locking usually requires that the gain medium has a large $3^{rd}$-order nonlinear coefficient. Moreover, it also demands critical cavity alignment and external perturbations to start. These drawbacks have severely limited their widespread applications.

Recently, graphene, an atomically thin sheet of carbon atoms arranged in a hexagonal honeycomb lattice, has attracted great interest as an alternative saturable absorber for mode-locking lasers. Graphene as a saturable absorber has the characteristics of ultrafast recovery time [8, 9], low saturation intensity, low cost, and easy fabrication. Most importantly, it also has wavelength independent saturable absorption over a broad spectral range from near-infrared to mid-infrared, which potentially allows the generation of even few optical-cycle pulses in the spectral region. Furthermore, the maximum absorption per layer of intrinsic graphene is ~ 2.3 %. This is especially suitable for the mode locking of bulk lasers as they have lower gain than the fiber lasers and cannot tolerate large non-saturable losses. The first graphene mode-locked fiber laser was demonstrated in 2009 [10, 11], and the first graphene mode locked bulk laser was a Nd:YAG ceramic laser. It generated transform-limited pulses with 4 ps duration centered at 1064 nm [12]. To date, graphene mode locking has been extensively investigated on various bulk lasers, mode locked pulses with duration ranging from picoseconds to femtoseconds have been demonstrated [12-24]. However, except for the case of a graphene mode-locked Cr:ZnS laser where 41 fs pulses centered at 2.4 μm [24] were generated, there have been no other reports on the generation of sub-100 fs pulses in graphene mode locked lasers. Although in the 1 μm wavelength region, sub-50 fs mode locked pulses have been obtained from various Yb-doped bulk lasers by using the SESAMs [25-28] and/or Kerr-lens mode locking technique [29, 30], the shortest graphene mode locked pulse in the Yb-doped bulk lasers was about 152 fs, generated form a Yb:YCOB bulk laser [19]. Recently, the Yb:CaYAlO$_4$ (Yb:CYA) crystal, another promising gain medium that has broad emission bandwidth and relatively high thermal conductivity, was developed [31]. Continuous-wave (CW) operation of the crystal has shown slope efficiency as high as 71% with an output power of 1.94 W. When mode-locked with a SESAM it has generated 156 fs pulses with 740 mW average output power [32]. The results suggest that it could be an excellent candidate for the ultrashort pulse generation.

In this letter we report on the generation of pulses as short as 30 fs (corresponding to 8.43 optical cycles) from a graphene mode locked Yb:CYA laser. The pulses are centered at 1068 nm with a spectral bandwidth of about 50 nm. To the best of our knowledge, this is the shortest pulse generated from the graphene mode locked lasers and the Yb-doped bulk lasers.

**2. Experimental setup**

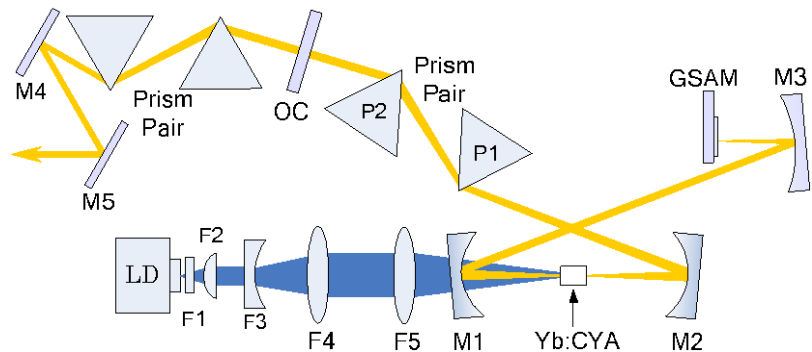

**Figure 1.** The schematic of the mode-locked laser setup based on the graphene saturable absorber mirror. F1: cylindrical lens; F2: aspherical lens; F3: concave lens; F4, F5: convex lens; M1, M2, M3: concave mirror; OC: output coupler; GSAM: graphene saturable absorber mirror; M4, M5: folding mirror.

The schematic of the mode-locked laser with a graphene saturable absorber mirror (GSAM) is shown in **Figure 1**. An Yb:CYA crystal with a Yb$^{3+}$ concentration of 8 at.% grown by the Czochralski method was used as the gain medium and positioned between two concave mirrors of M1 and M2. The crystal was a-cut and had a dimension of 3 mm in length and 3 × 3 mm$^2$ in cross-section. The sample was antireflection-coated for both the pump and laser wavelengths and was wrapped with indium foil, and tightly mounted in a water-cooled copper block whose temperature was water cooled to 18.0 ℃. To suppress the etalon effects and improve the stability of mode-locking, the crystal was placed a little bit tilted in the cavity. A distributed Bragg reflector (DBR) tapered diode laser at 980 nm was used in the experiment as the pump source. The pump light was collimated by a micro cylindrical lens (F1) with a focal length of about 0.5 mm and an aspheric lens (F2) with a focal length of about 3.5 mm. The collimated pump beam was further shaped by a telescope made up of a plano-concave lens F3 with the focal length of -25 mm and a plano-convex lens F4 with the focal length of 100 mm, and then focused into the crystal via a spherical lens (F5) of 80-mm focal length. The focused pump spot size was measured to be about 30 μm in diameter at the focal plane in the air. A standard X-folded cavity was adopted in the experiment for achieving simultaneously a suitable laser mode size in the crystal and on the GSAM. The two concave mirrors M1 and M2 had the same radius of curvature (ROC) of 50 mm. The laser beam was focused on the GSAM by a concave mirror M3 with a ROC of 100 mm. The three folding mirrors are dichroic mirrors. They were high reflection coated in the wavelength range from 1020 to 1200 nm and anti-reflection coated from 808 to 980 nm. Based on the ABCD matrix the waist radii of the laser beam inside the crystal and on the GSAM were calculated to be ~ 14 μm and ~ 60 μm, respectively. To reduce the losses of the cavity, a wedged plano-plano output coupler with a transmission of 0.4 % was used. A pair of SF10 prisms was used to compensate the group-delay dispersion (GDD) of the cavity. The prisms were placed in the arm containing the output coupler and had a tip-to-tip separation of ~ 40 cm, which could provide GDD of about – 1500 fs$^2$ per round trip.

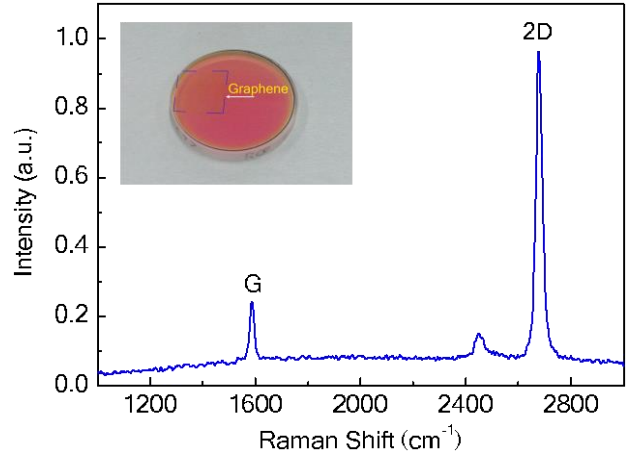

**Figure 2.** The Raman spectrum of GSAM excited by a 532 nm laser. Inset shows the image of the GSAM used in the experiment.

The GSAM used was fabricated by transferring a piece of high-quality monolayer graphene grown by chemical vapor deposition (CVD) method on a broadband high-reflection dielectric flat mirror. Comparing with other graphene synthesizing methods it has been shown that the CVD method could fabricate high-quality, large-area, homogeneous single layer graphene, which is good for the saturable absorber application. The graphene used in our experiment was grown on a 25 μm-thick Cu foil via the CVD method where a gas mixture of $CH_4$ and $H_2$ at 1000 ℃ was used [32]. The graphene was transferred onto the broadband reflective mirror by the same method reported in our previous papers [16, 19, 23]. The transferred graphene had a size of ∼ 1.0 × 1.0 $cm^2$, as shown in the inset of **Figure 2**. To assess the quality of graphene after the processes of transfer, we measured the Raman spectrum of the graphene on the mirror with a 532 nm laser source. Raman spectroscopy can offer a simple, fast, and non-destructive way to characterize the structure and layer number of graphene [34, 35]. The two characteristic Raman peaks of graphene designated as G and 2D can be clearly seen in Figure 2. The G peak is located at ∼ 1585 $cm^{-1}$ and 2D peak at ∼ 2676 $cm^{-1}$, and their full-width at half-maximum (FWHM) is approximately 19 $cm^{-1}$ and 32 $cm^{-1}$, respectively. The D peak, which is related to defects, was not observed at ∼ 1350 $cm^{-1}$ in the Raman signal. The indiscernible D peak and the high intensity ratio (∼ 4) of the 2D peak to G peak [34-36] confirm the presence of high-quality, single-layer graphene on the dielectric flat mirror. The linear reflectivity of the GSAM around 1 μm was measured to be about 95% with CW Yb:CYA laser in the experiment, which indicates there is little doping effect induced by substrate [19, 37]. In our previous works, as well as those of other research groups, the linear and nonlinear properties of the same type monolayer graphene at different wavelengths were investigated [15-20, 22]. Pump-probe experiments had revealed that the saturable absorption of the monolayer graphene had a similar feature that consists of an ultrafast recovery time of ~ 180 fs and a slower component of ~ 1.2 ps at different wavelengths [15, 17, 19]. The modulation depth and non-saturable loss were about ~ 0.6-0.7% and ~ 1.4-1.6% at around 1000 nm, respectively. And the saturation fluence was ~ 50 μJ/cm$^2$ at this wavelength region. All these parameters make graphene very suitable as a saturable absorber for bulk lasers to generate ultrashort pulses.

## 3. Results and discussion

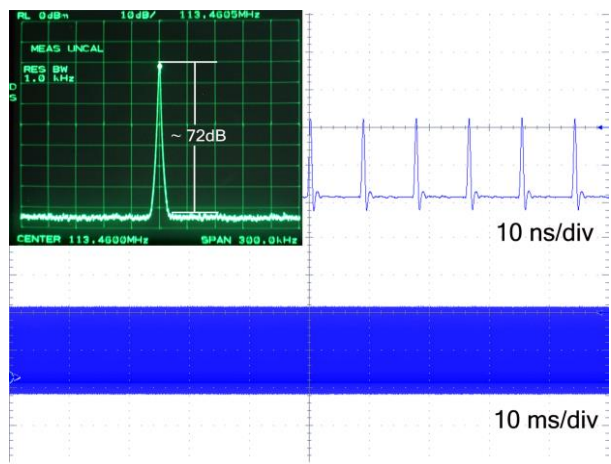

**Figure 3.** CW mode-locked pulse trains in nanosecond and millisecond time scales. Inset: RF spectrum around the fundamental repetition frequency.

In the experiment, stable CW mode locking could be easily initiated by the GSAM. Initially, the laser was in the CW regime. As the absorbed pump power increased to 1.17 W, the laser operation changed from CW operation directly to stable CW mode locking, and the corresponding average output power was 15.1 mW. No Q-switched mode-locking was observed. Under an incident pump power of ∼ 2.30 W the average output power reached a maximum value of 26.2 mW. When the pump power was further increased, the mode-locking became unstable, multiple pulses could be observed in the cavity and the mode locking eventually disappeared. But as the pump power was decreased to the previous values of CW mode locking, the mode-locking state could be re-established. **Figure 3** shows the typical CW mode-locked pulse trains in nanosecond and millisecond time scales monitored with a high-speed photoelectric detector (New Focus, 1611) and a digital oscilloscope with 1 GHz bandwidth (Tektronix, DPO7104). The corresponding radio frequency (RF) spectrum around the fundamental repetition frequency was measured with a resolution bandwidth (RBW) of 1 kHz, shown in the inset of Figure 3. The signal-to-noise ratio was as high as ∼ 72 dB, without any sign of the Q-switching instability. The pulse repetition rate was 113.5 MHz, corresponding to the laser cavity length of ∼ 1.32 m. The stable pulse trains and high signal-to-noise ratio evidenced the stable and clean CW mode locking. During the experiment, we tried to move the GSAM in the transverse direction to shift the laser spot out of the graphene region on the GSAM. Such a movement didn't change the cavity configuration. The laser then became CW emission and no sign of Kerr lens mode locking was observed. Based on the pulse energy in cavity and the calculated beam waist on the GSAM, we estimated the maximum fluence on the GSAM was about 510 μJ/cm$^2$. The mode-locking operation of the laser was stable and no damage was seen on the GSAM even the laser was operated under the maximum average output power.

**Figure 4** shows the optical spectrum of the mode-locked pulses measured by an optical spectrum analyzer (ANDO, AQ6315B) with the resolution of 0.5 nm. The optical spectrum is centered at 1068 nm and has a sech$^2$ profile with a (FWHM) bandwidth of ∼ 50 nm, which could support a Fourier-Transform-limited pulse of about 24 fs duration. The spatial laser beam profile measured by a laser beam profiler confirms the TEM$_{00}$ operation of the mode-locked pulses, as shown in the inset of Figure 4. It can be seen that the size of the output laser beam in the horizontal direction is a little bit wider than that in the vertical direction, which could be attributed to the broadening of the laser beam in the horizontal direction caused by the intracavity prism pair.

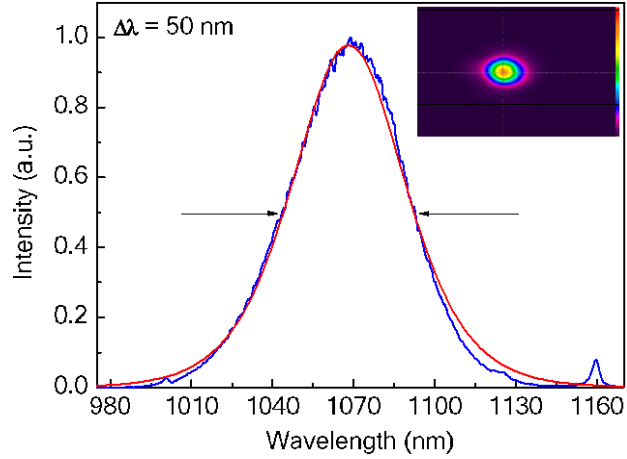

**Figure 4.** Optical spectrum of the mode-locked pulses. Inset: Spatial laser beam profile after the output coupler.

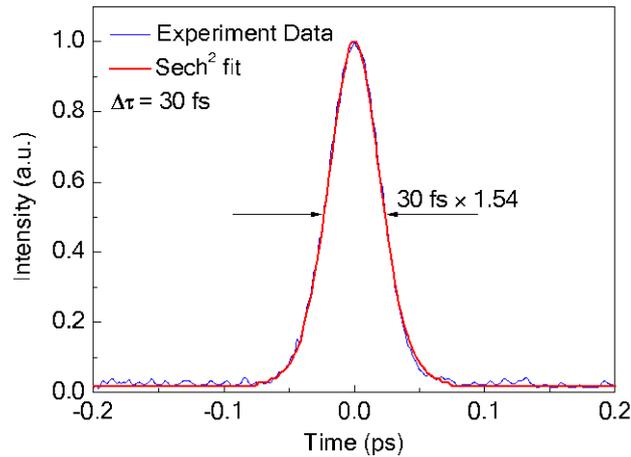

**Figure 5.** Autocorrelation trace of the mode-locked pulses after passing through the external compressor.

To eliminate the possible residual chirps and the slight spatial dispersion of the output pulses, a pair of SF10 prism was used outside the cavity (see Figure 1). After passing through the prism pair the mode-locked pulses was reflected by two folding mirrors before entering into the autocorrelator (APE, PulseCheck 50). With the specially ordered nonlinear crystal and a low dispersion beam splitter around the mode-locked center wavelength the autocorrelator could measure pulses as short as 30 fs with low measurement error (less than 1 %). By optimizing the insertion length of the prisms, the shortest pulses were obtained in the experiment. The final pulse duration was as short as 30 fs if a sech$^2$ pulse shape was assumed, as shown in **Figure 5**. The time-bandwidth product of the pulses was calculated to be about 0.39, which is close to that of the Fourier transform-limited pulses.

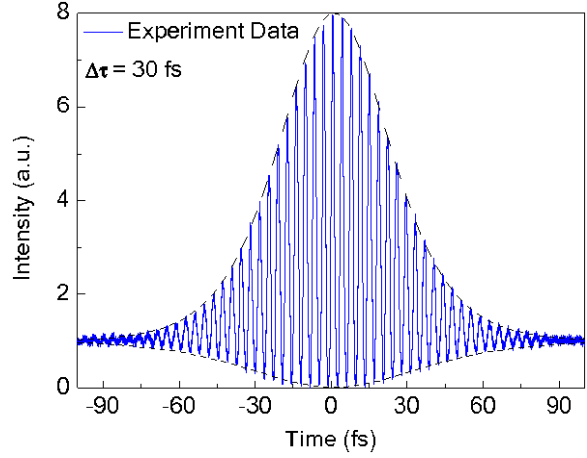

**Figure 6.** Interferometric autocorrelation trace of the mode-locked pulses. Dotted curves show the fitting envelopes assuming sech$^2$-shaped pulses.

To confirm the accuracy of measured pulse width, we also characterized the pulses with an interferometric autocorrelator (FEMTOCHROME, FR-103PD) because it was sensitive to chirps and may help us to extract more information about the pulses. **Figure 6** shows the measured interferometric autocorrelation trace of the mode-locked pulses. Assuming a sech$^2$-shaped pulse profile, the pulse duration was calculated again as about 30 fs, which well matches the result obtained from the intensity autocorrelation trace. In addition, the interferometric autocorrelation trace has exhibited fitted envelope profile that suggests no obvious chirp on the mode-locked pulses.

It was noticed that 32 fs pulses were generated from a Yb:CaGdAlO$_4$ (Yb:CGA) laser with the Kerr-lens mode locking technique recently [30]. The Yb:CGA and Yb:CYA belong to the same crystal family and have very similar mechanical, thermal and lasing properties. Comparing to the Kerr-lens mode-locked Yb:CGA laser, even shorter mode locked pulses have now been obtained from the Yb:CYA laser with a graphene saturable absorber. We note that due to the very small mode size in the laser crystal and the small output coupling used in our laser, the laser intensity at the position of laser crystal was about 312 GW/cm$^2$, which is much larger than that in Ref [30] (about 50 GW/cm$^2$). The very high intensity led to strong nonlinear self-phase modulation of the pulses propagating in the laser crystal. Therefore, we believe there was large nonlinear pulse shaping in the laser and the formed pulses are actually optical solitons. It is to point out that the average dynamics of a solitary pulse in a mode locked laser is determined by the complex Ginzburg-Landau equation. Different from the linear mode locking situation where in order to obtain the shortest pulses reducing the residual cavity dispersion is crucial, in the soliton operation regime the cavity dispersion can be balanced by the nonlinear phase modulation, and the soliton pulse width is determined by the mutual nonlinear interaction among the saturated gain and loss, gain dispersion, cavity dispersion and the nonlinear phase modulation. The soliton operation of our laser is further supported by the appearance of the Kelly sideband at around 1160 nm on the optical spectrum of the mode locked pulses. The Kelly sideband appeared only on one side of the optical spectrum, which suggests that the 2$^{nd}$-order dispersion is well compensated in the laser, and the formed soliton pulse width is mainly determined by the 3$^{rd}$-order dispersion of the cavity.

## 4. Conclusions

In conclusion, we have experimentally demonstrated 30 fs pulse generation in a Yb:CYA bulk laser using the graphene mode locking method. The mode locked pulses are centered at 1068 nm and have a spectral bandwidth of 50 nm. To the best of our knowledge, this is not only the first graphene mode-locked sub-100 fs pulses in the 1 μm wavelength region, but also the shortest pulses ever reported for graphene mode-locked bulk lasers and mode locked Yb-doped bulk lasers. The graphene mode-locking technique is easy to implement, comprising essentially a piece of high-quality, large-area single layer graphene and a broadband high reflection dielectric mirror to construct the graphene saturable absorber mirror (GSAM). Our experimental result shows that the mode locking performance of such a GSAM can be comparative with that of the Kerr lens mode locking. The experimental result demonstrated that the graphene mode locking could be a very promising practical technique for generating few-cycle optical pulses directly from a bulk lasers.

**Acknowledgements** The research is partially supported by the funds of Minister of Education (MOE) Singapore, under Grant No. 35/12, and also MOE Tier 1 grant "2-D Crystals as Platforms for Optoelectronics" R-143-000-556-112. We are grateful to Dr. Arne Heinrich from the Pantec Engineering AG for discussing and helping in beam shaping of diode laser, Dr. Katrin Paschke et al. from Ferdinand-Braun-Institut for supplying laser diode, and Mr. Lars Bucher, Mr. Edlef Büttner, et al. from APE GmbH for discussing the measurement of the mode-locked pulses.